\documentclass{iau}
\usepackage{graphicx}
\usepackage{epstopdf}

\title[IAUS 308.~~Evolution of steep-spectrum sources] %% give here short title %%
{Evolution of low-frequency contribution in emission of steep-spectrum radio sources}

\author[Alla P. Miroshnichenko]{Alla P. Miroshnichenko}   

\affiliation{Institute of Radio Astronomy of NAS of Ukraine, \\ 
UA-61002, Kharkov, Ukraine \\ email: {\tt mir@rian.kharkov.ua}}

\pubyear{2014}
\volume{308}  %% insert here IAU Symposium No.
\pagerange{xxx--xxx}
\jname{The Zeldovich Universe: Genesis and Growth of the Cosmic Web}
\editors{R. van de Weygaert, S. Shandarin, E. Saar \& J. Einasto, eds.}
\begin{document}

\maketitle

\begin{abstract}
We consider evolution properties of galaxies and quasars with steep radio spectrum 
at the decametre band from the UTR-2 catalogue. The ratios of source's monochromatic
luminosities at the decametre and high-frequency bands display the dependence on the
redshift, linear size, characteristic age of examined objects. At that, the mean
values of corresponding ratios for considered galaxies and quasars have enough close
quantities,testifying on the unified model of sources. We analyse obtained relations
for two types of steep-spectrum sources (with linear steep spectrum (S) and low-frequency
steepness after a break (C+)) from the UTR-2 catalogue.
\end{abstract}

\firstsection % if your document starts with a section,

\section{Introduction}

We continue to study the properties of the steep-spectrum sources from the Grakovo 
decametre survey (UTR-2 catalogue) within the frequency range 10 to 25 MHz (
\cite[Braude et al. 1978] {Braude_etal78}, 
\cite[Braude et al. 1979] {Braude_etal79},
\cite[Braude et al. 1981a]{Braude_etal81a}, 
\cite[Braude et al. 1981b]{Braude_etal81b}, 
\cite[Braude et al. 2003] {Braude_etal03}). 
This peculiar class of radio sources (the value of low-frequency spectral index
exceeds 1) corresponds to conception of the long evolution, when the critical frequency
of the synchrotron emission can displace to values less than 10 MHz. Befor (
\cite[Miroshnichenko 2012a] {Miroshnichenko12a},
\cite[Miroshnichenko 2012b] {Miroshnichenko12b},
\cite[Miroshnichenko 2013] {Miroshnichenko13})
we received estimates of the main physical parameters of quasars and galaxies 
with steep radio spectrum over the sample of objects at the decameter band 
(at the frame of the Lambda-CDM model of the Universe). The sample of objects with 
linear steep spectrum (type S) includes 78 galaxies and 55 quasars with flux 
density more than 10 Jy at the frequency 25 MHz. The sample of objects with break
steep spectrum (type C+) contains 52 galaxies and 36 quasars with flux density  
more than 10 Jy at the frequency 25 MHz. The optical and high-frequency data for 
examined sources have been got from the NED database. The redshift range of 
objects forms 0.017-3.570. Note, our calculations show that galaxies and quasars
with steep low-frequency spectra have the great luminosity (by order of $10^{28}$ 
W/(Hz ster) at the frequency 25 MHz) and very extended radio structure with linear size 
by order of 1 Mpc, and characteristic age by order of 100 million years ( 
\cite[Miroshnichenko 2012a] {Miroshnichenko12a},
\cite[Miroshnichenko 2012b] {Miroshnichenko12b},
\cite[Miroshnichenko 2013] {Miroshnichenko13}).

\section{Contribution of decametre emission in sources with S-type and C+ - type of steep spectra}

It is known that the decametre  emission of galaxies and quasars corresponds to 
emission of their extended regions, outlying from source's core. At the same time, 
the source's emission at the higher frequencies, mainly, is connected with the 
emission from the central region of radio source. The ratio of low-frequency
and high-frequency luminosities may be as characteristic of different substructures 
of objects indicating some features of their evolution. Moreover, one is not influenced
by the Universe model. So, we determine the ratios of the flux densities of emission
in the different bands: decametre (25 MHz), 
centimetre (5000 MHz), infrared (IR, K-band), optical (opt, V),X-ray (1 keV) band for quasars and galaxies from the steep-spectrum sample. These are identical to the ratios of the corresponding monochromatic luminosities (Table \ref{tab1}).
It is important that the mean values of the corresponding luminosity's ratios for 
considered quasars and galaxies in Table \ref{tab1} have enough close quantities.
Thus, the obtained characteristics of sources with steep radio spectrum are in concordance 
with the unified model of sources.

\begin{table}
  \begin{center}
  \caption{Mean values of the ratios of monochromatic luminosities at the different bands for quasars
  and galaxies with steep spectrum.}
  \label{tab1}

  \begin{tabular}{|l|c|c|}\hline 
{\bf Mean value of ratio} & {\bf Quasars} & {\bf Galaxies} 
    \\ \hline
 $<lg(S_25/S_5000)>$ & $1.69 \pm 0.08$ & $1.74 \pm 0.05$\\ \hline
 
 $<lg(S_25/S_IR)>$ & $4.30 \pm 0.11$ & $3.67 \pm 0.19$)\\ \hline
  
 $<lg(S_25/S_opt)>$ & $5.00 \pm 0.10$ & $5.15 \pm 0.12$\\ \hline
 
 $<lg(S_25/S_X)>$ & $7.78 \pm 0.17$ & $7.89 \pm 0.33$\\ \hline
 
 $<lg(S_IR/S_X)>$ & $3.54 \pm 0.20$ & $4.68 \pm 0.47$\\ \hline

  \end{tabular}

 \end{center}

\end{table}

\begin{figure}[h]
% \vspace*{-2.0 cm}
\begin{center}
\begin{minipage}[h]{0.45\linewidth}
 \includegraphics[width=1\linewidth]{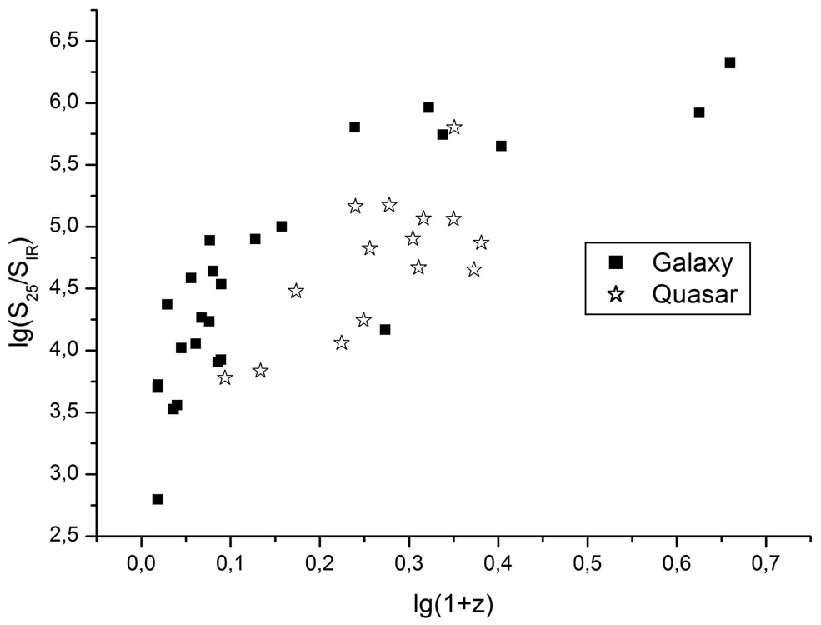} 
% \vspace*{-1.0 cm}
 \caption{The ratio of monochromatic luminosities of examined sources at the decametre 
 and infrared bands versus the redshift (for type S).}
 \label{fig1}
\end{minipage}
\hfill 
\begin{minipage}[h]{0.45\linewidth}
 \includegraphics[width=1\linewidth]{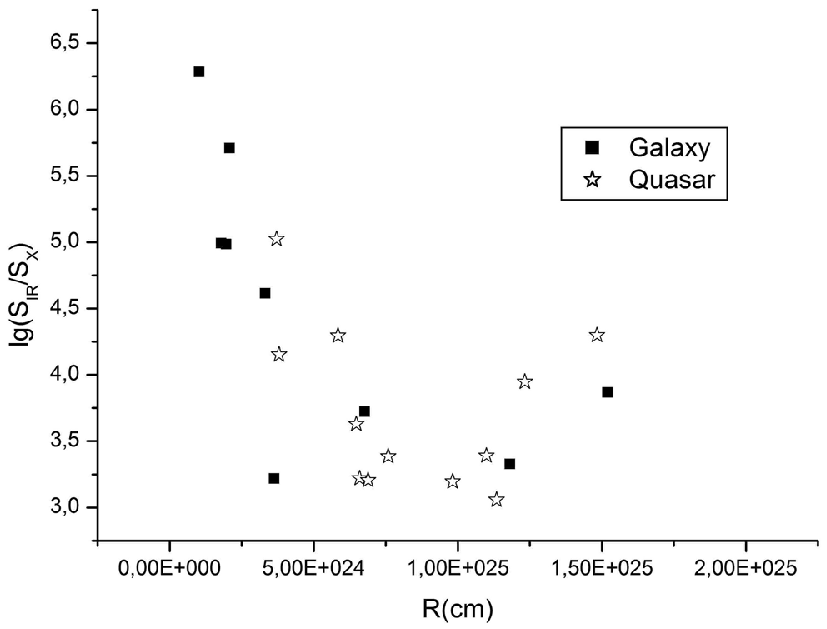} 
% \vspace*{-1.0 cm}
 \caption{The ratio of monochromatic luminosities of examined sources at the infrared
  and X-ray bands versus the linear size (for type S).}
 \label{fig2}
\end{minipage}
\end{center}
\end{figure}

\begin{figure}[h]
% \vspace*{-2.0 cm}
\begin{center}
\begin{minipage}[h]{0.45\linewidth}
 \includegraphics[width=1\linewidth]{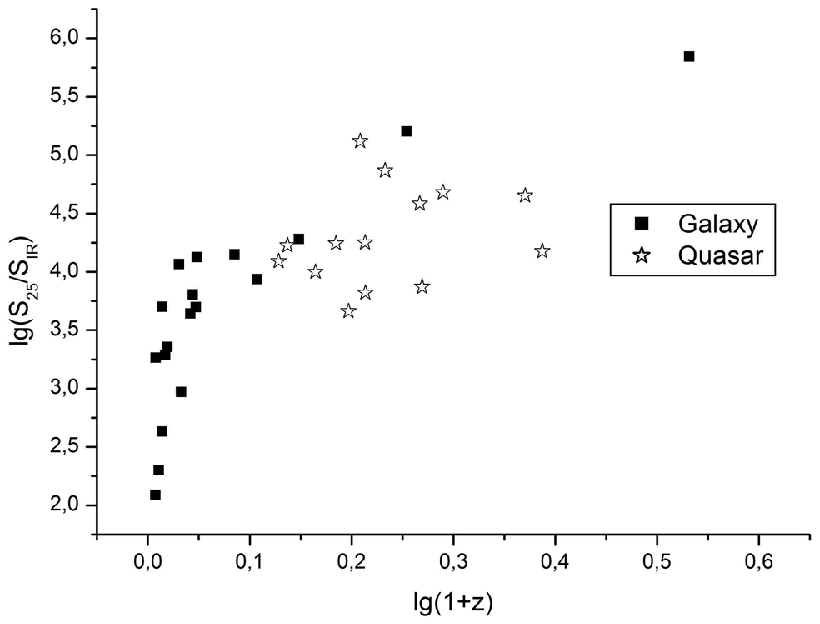} 
% \vspace*{-1.0 cm}
 \caption{The ratio of monochromatic luminosities of examined sources at the decametre
 and infrared bands versus the redshift (for type C+).}
   \label{fig3}
\end{minipage}
\hfill 
\begin{minipage}[h]{0.45\linewidth}
 \includegraphics[width=1\linewidth]{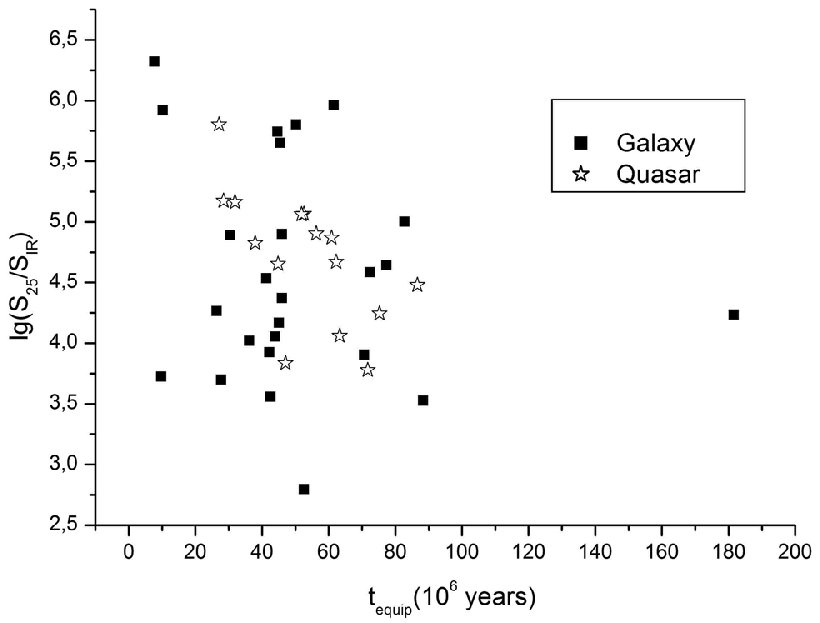} 
% \vspace*{-1.0 cm}
 \caption{The ratio of monochromatic luminosities of examined sources at the decametre 
 and infrared bands versus the characteristic age (for type S).}
   \label{fig4}
\end{minipage}
\end{center}
\end{figure}

We have received the relations for derived luminosity ratios of quasars and galaxies
with spectrum S and C+ versus the redshift, linear size, characteristic age 
(see Fig.\ref{fig1}-\ref{fig4}). These relations evidence for the essential cosmological 
evolution of luminosities of steep-spectrum sources. The interesting picture is displayed 
in the relation of infrared and X-ray luminosities versus the linear size of sample 
objects (Fig. \ref{fig2}). The founded two branches in this relation may testify on
the recurrence of the nucleus activity in galaxies and quasars with steep radio spectrum.  
  
%\begin{figure}[h]
%% \vspace*{-2.0 cm}
%\begin{center}
% \includegraphics[width=3.8in]{Mirosh_fig1.eps} 
%% \vspace*{-1.0 cm}
% \caption{The ratio of monochromatic luminosities of examined sources at the decametre 
% and infrared bands versus the redshift (for type S).}
%   \label{fig1}
%\end{center}
%\end{figure}
%
%\begin{figure}[h]
%% \vspace*{-2.0 cm}
%\begin{center}
% \includegraphics[width=3.8in]{Mirosh_fig2.eps} 
%% \vspace*{-1.0 cm}
% \caption{The ratio of monochromatic luminosities of examined sources at the infrared
%  and X-ray bands versus the linear size (for type S).}
%   \label{fig2}
%\end{center}
%\end{figure}
%
%\begin{figure}[h]
%% \vspace*{-2.0 cm}
%\begin{center}
% \includegraphics[width=3.8in]{Mirosh_fig3.eps} 
%% \vspace*{-1.0 cm}
% \caption{The ratio of monochromatic luminosities of examined sources at the decametre
% and infrared bands versus the redshift (for type C+).}
%   \label{fig3}
%\end{center}
%\end{figure}
%
%\begin{figure}[h]
%% \vspace*{-2.0 cm}
%\begin{center}
% \includegraphics[width=3.8in]{Mirosh_fig4.eps} 
%% \vspace*{-1.0 cm}
% \caption{The ratio of monochromatic luminosities of examined sources at the decametre 
% and infrared bands versus the characteristic age (for type S).}
%   \label{fig4}
%\end{center}
%\end{figure}

\section{Conclusions}

Galaxies and quasars with steep radio spectrum display the essential cosmological
evolution.
The relative contribution of the decametre emission in steep-spectrum sources increases
for more extended objects.
The revealed two branches in relation of the ratio of infrared and X-ray luminosity
versus the linear size of steep-spectrum galaxies and quasars may indicate on the
activity recurrence of sources.
Mutual similarity of the structure and the physical parameters of steep-spectrum
galaxies and quasars corresponds to the unified model of sources.


\begin{thebibliography}{}

\bibitem[Braude \etal\ (1978)]{Braude_etal78}
{Braude, S., Megn, A., Rashkovski, S., Ryabov, B. et al.} 1978,
\textit{Ap \& SS}, 54, 37 

\bibitem[Braude \etal\ (1979)]{Braude_etal79}
{Braude, S., Megn, A., Sokolov, K., Tkachenko, A., \& Sharykin, N.} 1979,
\textit{Ap \& SS}, 64, 73

\bibitem[Braude \etal\ (1981a)]{Braude_etal81a}
{Braude, S., Miroshnichenko, A., Sokolov, K., \& Sharykin, N.} 1981a,
\textit{Ap \& SS}, 74, 409
 
\bibitem[Braude \etal\ (1981b)]{Braude_etal81b}
{Braude, S., Miroshnichenko, A., Sokolov, K., \& Sharykin, N.} 1981b,
\textit{Ap \& SS}, 76, 279 

\bibitem[Braude \etal\ (2003)]{Braude_etal03}
{Braude, S., Miroshnichenko, A., Rashkovski, S., Sidorchuk, K. et al.} 2003,
\textit{Kinematics \& Physics of Celestial Bodies}, 19, 291
  
\bibitem[Miroshnichenko (2012a)]{Miroshnichenko12a}
{Miroshnichenko, A.} 2012a, 
\textit{Radio Physics \& Radio Astronomy}, 3, 215

\bibitem[Miroshnichenko (2012b)]{Miroshnichenko12b}
{Miroshnichenko, A.} 2012b, 
\textit{Odessa Astronomical Publications}, 25, 197

\bibitem[Miroshnichenko (2013)]{Miroshnichenko13}
{Miroshnichenko, A.} 2013, 
\textit{Odessa Astronomical Publications}, 26, 248


\end{thebibliography}
\end{document}